\documentclass[useAMS,usenatbib,a4]{mn2e}

\usepackage{times}
\usepackage{graphicx,bm,amssymb}
\usepackage{epsfig}
\usepackage{natbib}
\newcommand{\msun}{\mbox{$M_\odot$}}
\newcommand{\rsun}{\mbox{$R_\odot$}}
\def\be{\begin{eqnarray}}
\def\ee{\end{eqnarray}}
\def\lsim{\mathrel{\rlap{\lower3pt\hbox{\hskip1pt$\sim$}}
     \raise1pt\hbox{$<$}}} 
\def\gsim{\mathrel{\rlap{\lower3pt\hbox{\hskip1pt$\sim$}}
     \raise1pt\hbox{$>$}}} 

\begin{document}

\title[Hypercritical accretion in Kerr BHs with massive companions]{The need for hypercritical accretion in massive black-hole binaries with large Kerr parameters}

\author[Enrique Moreno M\'endez]{Enrique Moreno M\'endez\thanks{E-mail: emoreno@astro.uni-bonn.de} \\
Argelander-Institut f\"ur Astronomie, Bonn University, Auf dem H\"ugel 71, 53121 Bonn, Germany}

\maketitle

\begin{abstract}

Recent measurements of the Kerr parameters of the black holes in M33~X$-$7 and LMC~X$-$1 yield $a_\star\!=\!0.84\pm0.05$ and $a_\star\!=\!0.90^{+.04}_{-.09}$ respectively.
We study massive binary evolution scenarios that can reproduce such high values for the Kerr parameters.
We first discuss a model with Case C mass transfer leading to a common envelope and tidal synchronization of the primary before it collapses into a black hole.
We also study a Case M evolution model (which involves tidally-locked, rotationally-mixed, chemically-homogeneous stars in a close binary).
Our analysis suggests that, regardless of the specific scenario, the observed Kerr parameters for the black holes in M33~X$-$7 and LMC~X$-$1 had to be obtained through hypercritical mass accretion.

\end{abstract}



\section{Introduction}\label{sec-Intro}

When a star accretes mass onto its surface, it releases energy in the form of photons.  These in turn regulate the accretion rate to what is known, in the spherically symmetric case (or Bondi accretion, \citealt{Bon51}), as the Eddington limit:
\be
\dot{M}_{Edd}=\frac{L_{Edd}}{\epsilon c^2}=\frac{4\pi cR}{\epsilon\kappa_{es}},
~~~{\rm where }~~~
L_{Edd}=\frac{4\pi GMm_p c}{\sigma_T}
\ee
is the Eddington luminosity, with $m_p$ the mass of the proton, $\sigma_T$ the Thompson cross-section of the electron, $R$ and $M$ the radius and mass of the star and $\kappa_{es}$ the opacity of the infalling material, which is likely to be ionized hydrogen (so $\kappa\!=\!0.4$ cm$^2$ g$^{-1}$) and $\epsilon$ is the efficiency for converting mass into (photon) energy via the accretion process.  In general, if the accretion rate grows, the luminosity increases and self regulates the accretion rate to values below the Eddington limit. However, if the accretion rate grows to values which exceed this limit by a couple orders of magnitude, then photons become trapped.

\citet{Che81} and later \citet{Bro94} found that when the diffusion timescale of the photons generated by the accretion process is longer than the dynamical timescale of the accreting material, the photons become trapped and a shock forms (at $r_{sh}$) inside the photon-trapping radius ($r_{ph}$).
The shock diminishes the kinetic energy of the accreting material by a factor of $\sim50$ and converts it into thermal energy.
When the temperature reaches $T\sim1$ MeV, $e^-$, $e^+$ pairs are created.  These pairs annihilate into neutrino, anti-neutrino pairs
which can easily escape and allow the transport of energy out of the accreting BH.
This prevents the accretion luminosity from exceeding the photon $L_{Edd}$ as long as $r_{sh}<r_{ph}$.
However, $\dot{M}_{Edd}$ is exceeded by a factor of $\sim\!10^4$ \citep{Bro95}.
This is known in the literature as hypercritical mass accretion.

Mass accretion into a black hole (BH) instead of onto a neutron star (NS), such as in the cases calculated by \citet{Che81} and \citet{Bro94}, might make an even stronger case for hypercritical accretion.   This is because for a BH there is no surface onto which the infalling material can collide and radiate, but rather an event horizon through which matter passes uninhibitedly. Therefore the Eddington luminosity depends strictly on the efficiency to produce photons by the accreted material as it falls towards the event horizon.  On top of this, most scenarios, at least in binaries, do not involve spherical mass accretion but rather accretion through a disk, so the Eddington limit might not even be the most accurate prescription.

On the other hand, the material being accreted has angular momentum which must be lost before it can reach the event horizon of the BH.
Angular momentum loss from the accreting material is an important problem that will need to be addressed.
This is beyond the scope of the present work. Here we will only point out the need of hypercritical mass accretion in models which evolve massive binaries into the observed BH binaries, such as those of \citet{Lee02}, \citet{Mor07} or \citet{deM09}.

\citet{Lee02} and later \citet{Mor07} have modeled the evolution of 15 Galactic BH binaries.  They start from wide binaries (i.e., the initial orbital separation is  $a_i\gtrsim1,500~\rsun$) allowing the primary star in the binary to evolve as if it were a single star until it starts helium-shell burning.  At this point, the primary fills its Roche lobe and starts to transfer mass to the secondary.  Mass transfer during
He-shell burning is known in the literature as Case C mass transfer \citep[e.g.,][]{Heu94}.  The mass transfer to the less-massive secondary star shrinks the orbit until a common envelope sets in.  The secondary star spirals in while expelling the hydrogen envelope of the primary until the Roche-lobe overflow is stopped and the two stars orbit each other in a much tighter orbit (a few $\rsun$).  In such an orbit, and within a timescale of $10^2$ to $10^4$ years, the helium star becomes tidally synchronized before it collapses into a BH.  Therefore the spin periods $P_{spin}$ of the binary components coincide with the orbital period $P_{orb}$.

Assuming angular momentum is conserved during the collapse, knowing the orbital period allows a good estimate of the natal Kerr parameter of the BH, $a_\star\!=\!Jc/GM^2$ \citep[see, e.g., Fig.1 in][]{Bro07}, where $J$ is the angular momentum, and M is the mass of the collapsed object.
Here we will assume that the star is rotating as a solid body at the moment of collapse \footnote{It is usually accepted  that angular momentum is efficiently transported in the interiors of massive stars.  This is required to explain, for example, the slow spin rate of neutron stars~\citep{Heg05}.}.  This provides an upper limit to the available angular momentum, which translates into an upper limit to the natal Kerr parameter.

\citet{Lee02} and \citet{Mor07} estimated the Kerr parameters of 15 Galactic BH binaries. This was done by modeling the evolution of the orbital periods from their current to their pre explosion conditions.

Measurements of the Kerr parameters on several galactic BHs as well as LMC~X$-$1 and M33~X$-$7 have been performed by fitting the X-ray continuum.
This was done with a fully relativistic model of a thin disk around a Kerr BH whose plane is that of the binary and which assumes that the inner edge of the
disk is at the innermost stable circular orbit (ISCO).
These measurements rely squarely on the model of the disk and still need confirmation via other methods.
However, at present, they are the only available data and we will assume they have been determined meaningfully.

Among the measurements of the Kerr parameters of Galactic BHs there are those of GRO J1655$-$40 (XN Sco 94), 4U 1543$-$47 (Il Lupi)
\citep{Liu08,Gou09,Oro09,Sha06} and more recently, \citet{Ste10} have measured the Kerr parameter of XTE J1550$-$564, utilizing the aforementioned method as well as by modeling the Fe-K$\alpha$ line shape.
The match between the measurements of these three systems and the predictions suggests that the model of \citet{Lee02} and
\citet{Mor08}~\footnote{Case C mass transfer and  a  common envelope phase, followed by tidal locking and conservation of angular momentum during the stellar collapse.} represents a viable approach to study the formation and evolution of the Kerr parameter in black hole binaries.

\begin{table}
\begin{center}
\begin{tabular}{|c|c|c|c|c|}
\hline
    BH Binary      &        $M_{BH}$        &        $M_{sec}$       &  $P_{now}$ &          $a_\star$     \\
                   &       $[\msun]$        &        $[\msun]$       &   [days]   &                        \\
\hline
\hline
      LMC~X$-$1    & $10.30\pm1.34$  & $30.62\pm3.22$ &    $3.91$  & $0.90^{+0.04}_{-0.09}$ \\
      M33~X$-$7    & $15.65\pm1.45$  &  $70.0\pm6.9$  &    $3.45$  &     $0.84\pm0.05$      \\
\hline
\end{tabular}
\end{center}
\caption{The observed masses for BH and secondary star, orbital period and Kerr parameter, $a_\star$, for LMC~X$-$1 and M33~X$-$7
respectively.}\label{Tab:Binaries}
\end{table}

Extending the model from the galactic binaries to the massive binary in M33~X$-$7, \citet{Mor08} concluded that the measured Kerr parameter \citep{Liu08}
in this system must have evolved from an initial low value, $a_\star\!<\!0.1$, to its present state, $a_\star\!\simeq\!0.84\pm0.05$, by accreting about $5~\msun$ hypercritically (in its 2 to 3 Myrs lifetime; see Table~\ref{Tab:Binaries} for the masses, orbital period and Kerr parameters).

This model overlooks the fact that, given the large masses of the components in M33~X$-$7, it is unlikely they will go through Case C mass transfer.  However, this model still provides an upper limit to the natal Kerr parameter of the BH.  If late Case B (or any earlier mass transfer for that case) were to occur, a BH were still formed and the binary were to survive a merger, then the tidal locking might still occur.   However, mass-loss during the late stages of the primary would cause angular momentum from the star and the orbit to be lost, therefore reducing the expected Kerr parameter of the BH as compared to that for Case C mass transfer.

\begin{table}
\begin{center}
\begin{tabular}{|l|l|l|l|}
\hline
Symbol          &       Mass                        & Symbol          &       Period                 \\
\hline \hline
$M_{pri}$       & of the primary                    & $P_{orb}$       & Orbital                      \\
$M_{sec}$       & of the secondary                  & $P_{pri}$       & Spin of primary              \\
$M_{ZAMS}$      & at Zero-Age Main Seq.             & $P_{sec}$       & Spin of secondary            \\
$M_{BH}$        & of the BH                         & $P_{1}$         & Orbital, Pre-SN              \\
$M_{1}$         & Pre-SN                            & $P_{2}$         & Orbital, Post-SN             \\
$M_{2}$         & Post-SN                           & $P_{now}$       & Orbital, Observed            \\
$M_{now}$       & Observed                          & $P_{c}$         & Critical spin                \\
$M_{He}$        & Onset of He burning               & $P_{rm}$        & Minimum spin                 \\
$M_{acc}$       & Accreted                          &                 & for case M                   \\
$\Delta M$      & lost during SN                    &                 &                              \\
\hline
$M_{pri,ZAMS}$  & Primary at ZAMS                   & $P_{orb,2}$     &  Post-SN period              \\
\hline
\end{tabular}
\end{center}
\caption{Notation used throughout the text.  The lower two entries are examples of how subindices may be combined.}\label{Tab:Notation}
\end{table}

It is  important to point out that in the scenario for M33~X$-$7, as well as in those which will be discussed for LMC~X$-$1, mass transfer from the donor to the BH cannot occur through Roche-lobe overflow (RLOF).  This mode of mass transfer is unstable for large mass ratios~\citep[currently $q\!\simeq\!5$ and $q\!\simeq\!3$ respectively, where $q\!=\!M_{sec}/M_{BH}$, $M_{sec}$ is the mass of the secondary star and $M_{BH}$ that of the BH, as outlined in Table~\ref{Tab:Notation} ; see e.g. ][and references therein.]{Pod92}.
The instability arises from the fact that during RLOF the Roche lobe (RL) of the donor shrinks faster (in a dynamical timescale) than the star (which
responds in a thermal timescale).
Hence, as the RL shrinks more mass is transferred resulting in a further decrease of the orbital separation and RL.  This results in a runaway process leading to the formation of a common envelope, a spiral-in phase and a merger \citep[see][for a discussion on the subject]{Tau06}.

A possible mode of mass transfer that may avoid a merger is through stellar winds.  \citet{Moh07} have shown that wind mass transfer can be highly efficient.  
In their {\it wind-RLOF} scenario the wind fills the RL, is focused and channeled onto the accretor through the inner Lagrangian point L1.
The accretion rate can be as large as 70\%, as long as the wind velocity is less than the escape velocity from the RL surface\footnote{It is interesting to note that in the case of wind mass loss from the donor with $\alpha\!\simeq\!1/3$, $\beta\!=\!\delta\!=\!0$ in eq. 16.20 of \citet{Tau06} (where $\alpha$, $\beta$ and $\gamma$ represent the fractions of mass loss due to wind of the donor, ejection from the accreting BH and from a coplanar circumbinary disk, respectively) a mass-transfer efficiency $\epsilon\!=\!1-\alpha\!\simeq\!2/3$ would leave the orbital separation constant}.
This is the case when the system is in a tight orbit and the wind is still accelerating when it reaches the RL surface.  
The donor stars in M33~X$-$7 and LMC~X$-$1 are close to filling their RL \citep{Oro07,Oro09}, making this scenario a possibility. 
Interestingly, evidence for such a mode of mass transfer comes from observations of another BH binary system, Cyg X$-$1.  The massive donor is not filling its RL, but a component of the wind appears to be focused onto the BH  \citep{Sow98}.
Contrary to the RLOF case, wind RLOF is stable regardless of the value of $q$ because the wind rather than the star fills the RL, thus leaving the star hydrostatically stable and confined to its RL.
For donor stars as massive as those in M33~X$-$7 and LMC~X$-$1, the wind mass-loss rate is a few solar masses per Myr
($\sim10^{-6}~\msun$ yr$^{-1}$), large enough to transfer several solar masses during the main sequence lifetime of the donor.

In section~\ref{sec-CaseC} we discuss, as an example of the procedure with Case C mass transfer, LMC~X$-$1.  We first work through the model without including hypercritical accretion and show that it is not possible to obtain the observed $a_\star$ and orbital period.  We later include hypercritical accretion.

\citet{deM09} have proposed an alternative scenario to form BH binaries with massive companions that undergo quasi-chemically homogeneous
evolution~\citep[see, e.g., ][]{Yoo06}.
We consider this binary evolutionary path as well.
In section~\ref{sec-CaseM} we show that Case M scenario (tidally-locked, rotationally-mixed, chemically-homogeneous stars) also requires a phase of hypercritical accretion to explain the measurements of the Kerr parameters in M33~X$-$7 and in LMC~X$-$1.
We discuss our results, and the consequences of the relaxation of our assumptions on our results in section~\ref{sec-Disc}.
We show our conclusions in section~\ref{sec-Concl}.


\section{Case-C-Mass-Transfer Binary Evolution}\label{sec-CaseC}
\subsection{LMC~X$-$1 Without Hypercritical Accretion}\label{subsec-LMCX1woHA}

The present day Kerr parameter of the $10.30~\msun$ BH in LMC~X$-$1 is $a_\star\!=\!0.90$.  Its orbital period is 3.91 days, which means the distance between BH and companion is roughly $36~\rsun$.  This is slightly more than twice the $(17\pm0.8)~\rsun$ of the companion.
The companion is barely confined to its Roche lobe \citep[$R_{sec}\!\sim\!0.9R_{RL}$, with $R_{sec}$ the radius of the companion and $R_{RL}$ the Roche lobe radius, ][]{Oro09}.  Given that at present $M_{sec}\!\sim\!31~\msun$, we know this system can not be much older than $\sim\!5$ million years \citep{Gou09}.

If we do not allow hypercritical accretion the mass of the BH can not be considerably altered since its formation.  In fact the Eddington rate limits the accreted mass to the BH in these 5 Myrs to $M_{acc}\!<\!0.22~\msun$. As can be seen in fig. 6 of \citet{Bro00}, $0.22\msun$ cannot even produce an increment in $a_\star$ of $0.1$ on a $\sim\!10\msun$ BH.

The separation in the massive binary systems believed to be the progenitors of BH binary systems such as LMC~X$-$1 and M33~X$-$7 has to be quite small.
In particular the orbital separation is smallest after the common envelope phase, before the collapse of the primary, which means tides are very efficient at this point.  This is the reason why such massive binary systems are believed to be tidally synchronized when the BH forms \citep[e.g.,][]{vdH07}.

Assuming that no angular momentum is lost during the stellar-collapse phase, it is possible to calculate the upper limit of the natal Kerr parameter of the BH.
This is the procedure adopted by \citet{Lee02} to predict the Kerr parameters of GRO J1655$-$40 and 4U 1543$-$47, and here we adopt the same method.

A natal Kerr parameter such as $a_\star\!=\!0.90$ constrains the pre-explosion orbital period of the binary, $P_{1}$, to a value close to $0.3$ days \citep[see Fig.1 of][]{Bro07}, which for the current masses would imply an orbital separation smaller than $7~\rsun$.
Even assuming a tremendous amount of mass is lost during the explosion ($86~\msun$, which we will justify in the following discussion), this distance is not larger than $9.5~\rsun$.
Similarly to the scenario pictured for M33~X$-$7 in \citet{Mor08}, this is quite an unlikely situation.
In fact the two stars in such an orbit would undergo a common envelope phase and end up as a merger.
This is the strongest argument against the possibility that the observed Kerr parameter corresponds to the natal value.

Next, the formation of the BH must be such that the pre-explosion orbital period $P_{1}\!\sim\!0.3$ days is transformed to the post-explosion (present) orbital period $P_{2}\!=\!3.91$ days.   The mass loss in the Blaauw-Boersma (BB) explosion\footnote{When the primary star explodes, the binary loses mass asymmetrically with respect to its center of mass (CoM).  The CoM shifts towards the secondary.  The orbital velocity of the two stars is suddenly too large for the new mass of the system.  In order to conserve momentum the binary suffers a kick along the orbital plane.} \citep{Bla61,Boe61} is the only mechanism to achieve this since not much mass
(a few $\msun$ since the companion is an O7/O8 III star) is lost from the system afterwards.
Nevertheless, this is problematic.
As stated in \citet{Mor08}, if half the mass of the system is lost during the BB explosion the binary breaks apart~\citep[i.e., using $M_{BH}$ for the BH mass and $M_{sec}$ for the mass of the companion, $\Delta M\!=\!M_{BH}+M_{sec}$; see ][for a more detailed derivation]{Bro01}
\be
\left(\frac{P_{breakup}}{P_1}\right)=\left(1+\frac{\Delta M}{M_{BH}+M_{sec}}\right)^2=4. 
\label{eq:bu}\ee
Our scenario needs $P_2/P_1\!\sim\!10$ (or a mass loss of $\sim\!86~\msun$!), $P_{breakup}$, $P_1$ and $P_2$ representing binary break up, pre- and post-explosion periods.
As Eq.~\ref{eq:bu} shows, this can not be achieved during the explosion, as such a mass loss would break the system apart.

It is important to note that, given the geometry and the anticipated evolution of the binary system, we expect the rotational axes of both stars to be aligned with the orbital rotation axis.
This is because, due to the strong tides, the system is likely in a state of minimum energy.
In this state the spin axes have aligned, the orbits are circular and the spin period of the primary and secondary stars are tidally synchronized,
$P_{pri}\!=\!P_{sec}\!=\!P_{1}$ \citep[see, e.g.,][]{Zah77}.

The SN explosion of a rapidly rotating star will likely be cylindrically symmetric along the rotational axis, and thus a large (and temporally long with respect to the spin period) asymmetry in the collapse (e.g., a neutrino flux) will likely produce a kick along this axial direction \citep[see, e.g.,][]{Spr98}.  Hence, at least as a first order approximation, the SN kick will be perpendicular to the BB kick (and to the original orbital plane).  Thus the SN kick is not only unable to counteract the BB kick but actually makes the total kick larger with respect to the center of mass of the binary.

Therefore, the SN explosion must proceed with much less mass loss in order to prevent the system from breaking apart.  The SN must not expand the orbital period by more than $\sim\!4$ times its pre-SN value (i.e., 1.2 days).

The amount of mass loss (assuming a fast wind such that no mass is transferred to the BH) from the companion after the formation of the BH to expand the orbit from $P_2\!=\!1.2$ days to $P_{now}\!=\!3.91$ days can be obtained from
\be
\left(\frac{P_{now}}{P_2}\right)=\left(\frac{M_2}{M_{now}}\right)^2
\label{eq:ml}\ee
\citep[see][for a derivation]{Heu94}.  Using $M_{now}\!=\!31~\msun$ for the final mass, we obtain a post-explosion mass $M_2\!=\!56~\msun$
Given that a binary with two stars of ZAMS mass above $\gtrsim60~\msun$ does not fit in a 0.3-day orbit, and that the necessary mass loss to bring the orbit from 0.3 to 3.91 days is unrealistic, we must look for alternative evolutionary paths.


\subsection{LMC~X$-$1 With Hypercritical Accretion}\label{subsec-LMCX1wHA}

The present day period of the binary is $P_{now}\!=\!3.91$ days which corresponds to a Kerr parameter of $a_\star\!\sim\!0.1$ in Fig. 1 of \citet{Bro07}.  Mass transfer after the formation of the BH can only occur from the now more massive companion towards the BH.
Assuming conservative mass transfer\footnote{It is interesting to note that \citet{Oro09} estimate a ZAMS mass of $\sim\!35\msun$ for the secondary star, similar to what conservative mass transfer would require.  However, as discussed earlier, this must occur through wind RLOF.}
implies that the orbit could only shrink from the time of the collapse until the present day.  This means that $a_\star\!\sim\!0.1$ is an upper limit for the natal Kerr parameter of the BH.  Most of the spin must be acquired through hypercritical mass accretion, that is, the original BH mass has to increase between $50\%$ and $70\%$ (or some $4~\msun$ to $5~\msun$) from its natal value to acquire the observed $a_\star\!=\!0.90^{+0.04}_{-0.09}$ value (see Fig.6 in~\citealt{Bro00}) in less than 5 Myrs.

Assuming that the re-circularization of the orbit does not alter the orbital period considerably, and that there is conservative mass transfer from the companion to
the BH, we can reconstruct the post-explosion orbital period, $P_2$, starting from the current one, $P_{now}$, in LMC~X$-$1~\citep[see, e.g., eqs. 90-97 in][for a derivation]{Heu94}:
\be
P_2&=&\left(\frac{M_{BH,now}\times M_{sec,now}}{M_{BH,2}\times M_{sec,2}}\right)^3 P_{now} \nonumber \\
   &=&\left(\frac{10.3\msun\times 30.6\msun}{6.3\msun\times 34.6\msun}\right)^3 3.91~{\rm days} = 11.8~{\rm days,}
\ee\label{eq:mt}
where $4~\msun$ are transferred or $P_2\!=\!18.2$ days if $5~\msun$ are transferred (where $M_{BH}$ represents the BH mass, $M_{sec}$ is the mass of the secondary, and the subindices $_2$ and $_{now}$ refer to post-explosion and present values).
This restricts the natal $a_\star\!<\!0.05$.  Nevertheless, if there was mass lost during the formation of the BH, the pre-explosion period had to be
somewhat smaller but taking into account the limits imposed by eq.~\ref{eq:bu}.
Therefore, assuming half of the mass of the system was lost during the explosion, the lower limit on the orbital period is 3 to 4 days.
This translates to a maximum natal Kerr parameter $a_\star\!<\!0.15$.

We can argue that the BB explosion leading to the formation of the BH could not lose close to half the mass of the system when it was formed.
First of all, a BH with a mass  $\gtrsim6\msun$ was created.
This implies that the SN shock wave had to be stalled at least until most of this material was accreted, otherwise this material would be lost from the binary.
Eventually, the stalled shock subsided and the material was advected into the BH.
As the BH formed and grew there were at least two mechanisms which could eventually launch an explosion {\bf if} an accretion disk was able to form.
The first scenario is the Blandford-Znajek (BZ) mechanism \citep{Bla77}.  The second one involves neutrino pair annihilation~\citep[see, e.g.][]{Mac99}.
However a disk might never form in this scenario given that the available angular momentum in this scenario is rather small.

For the sake of obtaining an idea of the available energy for an explosion after the BH is formed, let us discuss the first scenario.
Using the formalism in \citet{Lee02} we obtain $E_{BZ}\!\sim\!15$ bethes ($15\times10^{51}$ ergs) for $M_{BH}\!=\!6~\msun$ with $a_\star\!\sim\!0.15$ but
only about 4 bethes if $a_\star\!\sim\!0.08$ (we obtain this number in the following paragraph).
These numbers are much lower than the hundreds of bethes available in the rotational energy of the Galactic BH binaries~\citep{Bro07}.
This amount of energy may not be enough to produce a powerful explosion as only a fraction of this energy will be kinetic.

Another point to keep in mind is that for the secondary star to be $M_{sec,now}\!\sim\!31~\msun$, one can estimate that we need
$35~\msun\!<\!M_{sec,ZAMS}\!<\!40~\msun$ given that the star has not filled its Roche lobe, assuming a lifetime of 5 Myr and a mass loss of $\sim\!10^{-6}~\msun$ yr$^{-1}$~\citep[also,][by measuring the luminosity and temperature of the companion, estimate $M_{sec,ZAMS}\!\sim\!35~\msun$]{Oro09}. But this implies that the primary might have been $40~\msun\!<\!M_{pri,ZAMS}\!<\!50~\msun$.
Using the empirical relation~\citep{Lee02}
\[M_{He}\!=\!0.08(M_{ZAMS}/\msun)^{1.45}\msun,\]
we can estimate that the mass of the
He star was about $M_{He}\!\sim\!20~\msun$.  So, forming a BH with $M_{BH}\gtrsim 6~\msun$ implies that, in eq.~\ref{eq:bu}, $\Delta M\!\sim\!14~\msun$.
Therefore, the minimum pre-explosion
\[P_{1}\!=\![1+14/(6+35)]^{-2}\times12~{\rm days} \!=\!6.7~{\rm days}\]
for this scenario.  This equates to a natal Kerr parameter of $a_\star\!\sim\!0.08$.



Similar to the case of M33~X$-$7 \citep{Mor08}, the available information on LMC~X$-$1 points to an explosion where little mass was lost when the BH was formed.
This supports the scenario of a pre- (and post-) explosion period roughly between 12 and 18 days, which has been shortened, by mass transfer to the BH,
down to the presently observed 3.91 days.
Such mass transfer had to be hypercritical, leading to a growth of the Kerr parameter from $a_\star\lesssim 0.1$ to its currently observed $0.9$ value.

\section{Case-M Binary Evolution}\label{sec-CaseM}

As proposed in \citet{deM09}, we now consider Case M evolution of massive binaries, i.e., tidally-locked, rotationally-mixed, chemically-homogeneous stars.
The radii of stars in such an evolutionary path barely change as they evolve.
In the following calculations we assume that the radii of the stars do not change substantially from their ZAMS radii
\footnote{We assume that the radii of Case-M stars do not change significatively during the whole evolution.
Research is underway to establish whether the relaxation of this assumption may produce larger Kerr parameters.}.
On one hand, this will prevent them from filling their Roche lobes.
On the other hand, we assume they do not contract enough after the main sequence to substantially alter their spin.

\subsection{Without Hypercritical Accretion}\label{subsec-RotMix}

\subsubsection{M33~X$-$7}\label{subsubsec-M33X7}

In M33~X$-$7 we presently observe a $M_{sec}\!\sim\!70~\msun$ companion orbiting a $M_{BH}\!\sim\!15~\msun$ BH in a $P_{now}\!\sim\!3.45$-day orbit \citep{Liu08}.
To form a massive binary with a Kerr parameter of $a_\star\!=\!0.84$ and such an orbital period the system has to lose a substantial amount of mass at different stages.  Next we reconstruct the path that requires the lowest amount of mass loss, i.e. the stars with the lowest ZAMS masses that can explain the observed masses and orbital period.

The binary orbits in $P_{orb,1}\!=\!0.4$ days.  This provides the observed $a_\star\!=\!0.84$~\citep[see figure 3 in ][]{BLMM08} of the $15~\msun$ BH.
During the formation of the BH, the primary losses almost half the mass of the system i.e., using eq.~\ref{eq:bu}, and the mass we estimate for the secondary on the following paragraph
\[\Delta M\sim M_{BH}+M_{sec,2} = 15~\msun+103~\msun = 118~\msun,\,\, {\rm so}\]
\[P_{orb,2} = [1+118~\msun/(15+103)~\msun]^2\times0.4~{\rm day} = 1.6~{\rm day}.\]
After the formation of the BH, the secondary loses $(M_{sec,2}-M_{sec,now})\!\sim\!33~\msun$ this changes the post-collapse $P_{orb,2}\!\sim\!1.6$ day orbital period to the present one (eq.~\ref{eq:ml}),
\[P_2 = (103~\msun/70~\msun)^2\times1.6~{\rm day} = 3.45~{\rm day}.\]
This implies that, prior to the formation of the BH, the primary star, with $M_{pri,1}\!=\!15~\msun+118~\msun\!=\!133\msun$, and the secondary, $M_{sec,1}\!=\!70~\msun+33~\msun\!=\!103~\msun$ must orbit within $\lesssim 14~\rsun$.
As stated above, these are the smallest masses necessary.
If we chose smaller $\Delta M$ during the collapse and a larger wind mass loss from the secondary, the estimate becomes larger.
It is evident that such stars do not fit in this orbit.  A common envelope would merge them extremely rapidly.
Notice that this scenario has not yet accounted for mass loss previous to the collapse of the primary into the BH.
This would mean the stars were more massive and closer at ZAMS.
From the above it is clear then that this is an unlikely scenario to reproduce the observed values of stellar masses, orbital period and Kerr parameter in M33~X$-$7.

\subsubsection{LMC~X$-$1}\label{subsubsec-LMCX1}

In the case of LMC~X$-$1, the situation could be even tighter as the observed Kerr parameter is closer to $a_\star\!\sim\!0.9$ or a pre-collapse period close to $P_{orb,1}\!\sim\!0.3$ days.
Then, following the procedure of subsection~\ref{subsubsec-M33X7}
\[\Delta M\sim M_{BH}+M_{sec,2} = 10~\msun+56~\msun = 66~\msun,\,\, {\rm so}\]
\[P_{orb,2} = [1+66~\msun/(10+56)~\msun]^2\times0.3~{\rm day} = 1.2~{\rm day}.\]
After the formation of the BH, the secondary loses $\sim\!25~\msun$ which changes the post-collapse $P_{orb,2}\!\sim\!1.2$ day orbital period to the present one (eq.~\ref{eq:ml}),
\[P_2 = (56~\msun/31~\msun)^2\times1.6~{\rm day} = 3.9~{\rm day}.\]
This implies that the primary star, with $M_{pri,1}\!=\!10~\msun+66~\msun\!=\!76~\msun$, and the secondary, $M_{sec,1}\!=\!31~\msun+25~\msun\!=\!56~\msun$ must orbit within $\lesssim 10~\rsun$.

Again, in this scenario the orbit is too tight to fit the two stars; from table~\ref{Tab:Rot} we obtain that the actual ($R_{ZAMS}$) radii for rotationally-mixed, chemically-homogeneous stars of mass $70~\msun$ are $\!>\!10~\rsun$ when the metallicity is around $Z\!\sim\!(0.2-0.1)Z_\odot$.
These radii are very close to the orbital separation.
This strongly suggests that even rotationally-mixed chemically-homogeneous stars could not orbit close enough to produce the large Kerr parameters at the time the BHs were formed without merging the two stars.


\subsection{Hypercritical Accretion}\label{subsec-RM-HA}

Perhaps one of the most relevant points to notice in favor of hypercritical accretion after Case~M binary evolution is that in order to achieve the
observed Kerr parameters one seems to need a spin period smaller than the critical-rotation period of such stars at ZAMS (see table~\ref{Tab:Rot}).
This is regardless whether the stars fit or not in their Roche lobes and their orbits.
Nevertheless, this difficulty may be overcome by the shrinking of the core after the exhaustion of H.
This, however, breaks the tidal lock with the companion, and unless the ratio of stellar radius to orbital separation changes substantially, tides (as well
as winds) will eventually remove the gain in spin.

If we instead allow hypercritical accretion to occur into the BH, we can avoid the discrepancy between the current orbital period and Kerr parameter.
Nonetheless, there are still important constraints to be addressed regarding the allowed orbits.
The most important for LMC~X$-$1 and M33~X$-$7 being the following two:
\begin{enumerate}
\item{$P_{orb}\!=\!P_{spin}\le P_{rm}$, the tidally-synchronized primary star must rotate faster than the minimum critical spin below which rotational mixing
is not efficient enough to keep the star compact.
Otherwise the star will expand, fill its Roche lobe, transfer mass, shrink the orbit and produce a common envelope and, most likely, end up in a merger.}
\item{The secondary star must be either a main sequence star or must also be in a Case-M evolutionary track.
Otherwise we end up, again, with a merger and a larger but unobservable Kerr BH.}
\end{enumerate}

Contrary to what was discussed in section~\ref{subsec-RotMix}, in this scenario the stars are not required to rotate with enough angular momentum to produce the observed Kerr parameters.
Instead, we demand approximately five solar masses to be transferred into the BHs (see fig. 6 in \citealt{Bro00}) in both, M33~X$-$7 and LMC~X$-$1.
Otherwise it is not possible to obtain at the same time the observed Kerr parameters and orbital periods.

\citet{Yoo06} (Fig. 3) note that the main requirement for rotational mixing inside a star is a spin period of at least $20\%$ to $35\%$ of its critical rotation period ($P_c$), i.e. where gravity on its equator is balanced by centrifugal acceleration.
In tidally synchronized stars this means the orbital period $P_{orb}\le P_{rm}\!\simeq\!5 P_c$ (where $P_{rm}$ is the period necessary in order to have rotational mixing in the star).
As we have mentioned above, this might be necessary for both stars.

\begin{table}
\begin{center}
\begin{tabular}{|c|c|c|c|c|c|}
\hline
  $M_{sec}$  &   Radius    &     $P_c$     &    $\%P_c$    &    $P_{rm}$  &  $P_{now}$ \\
    $[\msun]$& $[\rsun]$  &     [days]    &               &    [days]    &   [days]   \\
\hline
\hline
      30    &      7      &     $0.39$    &     $35$      &    $1.2$     &   $3.91$   \\
      70    &     10      &     $0.44$    &   $20-25$     &  $2.2-1.8$   &   $3.45$   \\
\hline
\end{tabular}
\end{center}
\caption{The critical spin period, $P_c$, below which stars of the given mass and radius start losing mass from the equator.  $P_{rm}$ is the minimum necessary spin period for rotational mixing to occur.  $P_{now}$ are the current orbital periods in LMC~X$-$1 and M33~X$-$7, respectively.  ZAMS Radii in the second column from Yoon S.-C. (2009, private communication) for stars with metallicity around $Z\!\sim\!(0.2-0.1)Z_\odot$.}\label{Tab:Rot}
\end{table}

From Table~\ref{Tab:Rot} it is clear that $P_{orb,now}\!>\!P_{rm}$ for the secondary stars in both, M33~X$-$7 and LMC~X$-$1, this means that they cannot be
currently tidally synchronized and rotationally mixed.  The future fate of these systems will be very different depending on whether the secondary stars
manage to stay in Case~M or become tidally synchronized.

The orbital separation of the binary for the (ZAMS) stars in M33~X$-$7 to be tidally synchronized to the point where rotational mixing occurs would
be between $33~\rsun$ and $38~\rsun$ (assuming $M_{sec}\!=\!75~\msun$ and $M_{pri}\!=\!85~\msun$), and for LMC~X$-$1 it would be $\sim\!20~\rsun$ (assuming
$M_{sec}\!=\!35~\msun$ and $M_{pri}\!=\!45~\msun$).
Comparing these values to those in table~\ref{Tab:Rot} shows that these stars would barely be within their Roche lobes \citep[using ][we obtain $R_{RL}\!\sim\!7.7~\rsun$ for LMC~X$-$1 with $P_{orb}\!=\!1.2$ days; $R_{RL}\!\sim\!14~\rsun$ for M33X$-$7 with $P_{orb}\!=\!2$ days]{Egg83}.

These scenarios seem to be extremely tight, especially because no mass loss has been considered yet, and that the stellar radii are very close to their Roche lobe radii.  However, it is possible that they will produce binaries like those in M33~X$-$7 or LMC~X$-$1.

Following these assumptions, and not accounting for the substantial mass and angular momentum loss during core-He burning, a period of 1.2 days would give
a maximum $a_\star\!<\!0.3$ for the BH in LMC~X$-$1.  A period of 2 days would produce a BH with a maximum $a_\star\!<\!0.2$ for M33~X$-$7.
Since in this scenario, like in Case C mass transfer, we still face the the problem of unstable mass transfer due to the high values of $q$, it is clear
that the observed Kerr parameters only seem attainable through a phase of wind RLOF with $\sim\!5~\msun$ accreted hypercritically by the BH.


\section{Discussion}\label{sec-Disc}

To explain the observed Kerr parameters in massive BH binaries, here we have restricted our study exclusively to Case~C mass transfer and Case~M binary
evolution without covering other possible evolutionary scenarios.

On one side of the initial-orbital-separation spectrum, Case~C mass transfer spins the BH progenitor at the latest possible time before the stellar collapse (compared to Case~A and B mass transfer).
This allows a much shorter time for the stars to lose mass and angular momentum by means of stellar winds and/or tidal interactions. Most of the angular momentum can be retained until stellar collapse and contribute to the BH spin.

On the other extreme,  i.e. for very short orbital periods, Case M spins the progenitor early in the evolution and maintains it tidally locked throughout most of its lifetime. Therefore, under the assumption that the stellar radii do not change too much during the evolution of chemically homogeneous stars, Case M appears as a natural candidate to produce fast rotating collapsing stars.

Alternatively, Case~A and B are scenarios in which mass transfer from the primary occurs early in the lifetime of the binary system.
This means that mass and angular momentum can be efficiently lost in later phases of stellar evolution.
Hence, a slowly rotating BH will be produced in such scenarios.

Therefore we consider the Kerr parameters estimated for Case~C and Case~M mass transfer to be upper limits.
Even if we have not studied Case~A and Case~B mass transfer scenarios in detail, it seems very difficult to explain the observed Kerr parameters avoiding a hypercritical accretion phase.

In principle one may also suggest the interaction of stars in a triple or even multiple system.
Perhaps the most difficult part to justify in a triple scenario is how to bring in the BH and the massive secondary into the currently observed tight orbit.
After Case C mass transfer (where we have shown that the angular momentum is less likely to be lost), when the primary transfers mass and produces a
common envelope which allows the first companion to spiral in and merge the price to pay is the loss of the envelope of the primary \citep[see, e.g., the discussion in ][]{Lee02}.
Therefore, narrowing the orbital separation of the massive secondary may be difficult.
A likely scenario would be to wait for the massive secondary to expand, transfer mass to the BH and allow it to spiral in to a shorter orbital period.
However, neither LMC~X$-$1 nor M33~X$-$7, seem to have such evolved companion, in fact these seem to be still in the main sequence and within their Roche lobes.
In general, in the case of a multiple system, finding a way to reproduce the observed separation and Kerr parameter becomes a very complex exercise, and it  is beyond what we can do at present.

It is also interesting to speculate what the final fate of these systems will be if they have evolved through Case~M.
The fact that they can no longer be both, tidally locked and in Case~M produces very different outcomes.
After the SN and further wind mass loss from the secondary, the orbit has grown larger making tidal locking less efficient, if the synchronization timescale
is longer than the remaining lifetime of the massive secondaries we could still expect them to be rotating rapidly.
If the spin periods stay below $P_{rm}$, they will likely stay in Case~M.
If the second SN does not disrupt the binaries they will eventually form BH-BH (or BH-NS if enough mass is lost during the remaining evolution) binaries.
Like for the first-born BH, the natal Kerr parameters should be small, $a_\star<0.3$ for LMC~X$-$1 and $a_\star<0.2$ for M33~X$-$7.
Nevertheless these will not be able to increase given the lack of a mass donor.
If this is not the case and the secondary stars are being spun down by tidal synchronization, they will eventually fill their Roche lobes before the end of
their main sequence.  This will transfer mass to the BHs and narrow the orbits.
Their fate will be a merger with their respective BHs.
Given the large Kerr parameters of the BHs in these two systems, it could be expected that such an event will produce long gamma-ray bursts.
In the end, a massive but non-observable Kerr BH will be the remnant.

Recently~\citet{Val10} have suggested a formation mechanism for M33~X$-$7 which involves Case~A mass transfer.
Their model does not include rotational mixing, however it is compared with that of \citet{Hir05} which does include it in order to use the stellar angular momentum at the end of the main sequence calculated in this other model.  The result nevertheless should be significantly different.
Furthermore, as we have previously pointed out, most of the mass and angular momentum loss occurs at letter stages.  More important still, they obtain a system where the WR star is synchronized with the massive companion in an orbital period longer than 3 days.  Hence, the Kerr parameter of the resulting BH would not be larger than 0.1.  This, contrary to what is claimed in the paper, is not consistent with the observed value.  Hence, this model would still need wind RLOF and hypercritical accretion to explain the observed Kerr parameter.

As this paper prepares for press, a new paper, \citet{Axe10} has appeared on the spin parameter of Cygnus X$-$1.  The authors reach similar results as those from this paper, however, they conclude that the mechanism of \citet{Blo07} to spin NSs up through a stationary accretion shock instability (SASI) can produce the measured $a_\star\!=\!0.48\pm0.01$.  In a future paper we will demonstrate that this mechanism would be extremely difficult to justify for BHs with massive companions such as M33 X$-$7, LMC X$-$1 or Cyg X$-$1 because of conservation laws.


\section{Conclusions}\label{sec-Concl}

LMC~X$-$1 and M33~X$-$7 are two BH binaries where the masses of the components, the orbital period and the spin of the BH have been well constrained
observationally.  We have shown that, if we model the evolution of the progenitors of such systems with the standard methods, we cannot explain the current state of affairs in such systems.

We have discussed that Case-C-mass-transfer binary evolution provides an upper limit to the attainable natal spin of the BH (over Case~A and Case~B).
By stretching this model to its limits, we are able to estimate an upper limit to the Kerr parameters of LMC~X$-$1 and M33~X$-$7 of $a_\star<0.15$, well below the observed values of 0.90 and 0.84, respectively.

We have further attempted to obtain an evolutionary model which may explain the observations through Case~M evolution.  This model requires a very finely tuned system in order to reproduce the observed BH binaries but does seem to provide the largest possible Kerr parameters ($a_\star<0.3$ for LMC~X$-$1 and $a_\star<0.2$ for M33 X$-$8).
However, like the rest of the models, Case~M fails to come close to the observed values.
It is interesting to study whether relaxing some of the assumptions on this model might provide larger Kerr parameters, but one must remember that we still have not considered mass nor angular momentum loss at late evolutionary stages.  In the end, this model may still suffer the fate of Case~A and Case~B where the mass loss during He burning removes the outer layers of the star and with them, most of the angular momentum necessary to produce a substantial Kerr
parameter.

Both, Case-C-mass-transfer and Case-M models for binary evolution suggest that hypercritical accretion is necessary in order to explain the
observations of the Kerr parameters of the BHs in LMC~X$-$1 and M33~X$-$7.  Nevertheless, given the current mass ratios, this implies that efficient wind
mass transfer between the companion and the BH must occur.

\section*{Acknowledgements}
The author would like to thank the referee, M. Cantiello, S.E. de Mink, N. Langer, S.-C. Yoon, T. Tauris, S. Mohamed, H. Neilson, T. Decressin, J.S. Vink and Ph. Podsiadlowski for useful comments, discussions and insight.


\clearpage

\begin{thebibliography}{99}


\bibitem[\protect\citeauthoryear{Axelson et al.}{2010}]{Axe10}
Axelsson, M., Church, R.P., Davies, M.B., Levan, A.J. \& Ryde, F., 2010, MNRAS Accepted, arXiv:1011.4528v1.

\bibitem[\protect\citeauthoryear{Blaauw}{1961}]{Bla61}
Blaauw A. 1961, BAN 15, 265.

\bibitem[\protect\citeauthoryear{Blandford \& Znajek}{1977}]{Bla77}
Blandford R.D. and Znajek R.L. 1977, MNRAS 179, 433.

\bibitem[\protect\citeauthoryear{Blondin \& Mezzacappa}{2007}]{Blo07}
Blondin, J.M. \& Mezzacappa, A. 2007, Nature 445, 58.

\bibitem[\protect\citeauthoryear{Boersma}{1961}]{Boe61}
Boersma J. 1961, BAN 15, 291.

\bibitem[\protect\citeauthoryear{Bondi}{1951}]{Bon51}
Bondi H. 1952, MNRAS 112, 195.

\bibitem[\protect\citeauthoryear{Brown \& Weingartner}{1994}]{Bro94}
Brown G.E. and Weingartner J.C. 1994, ApJ 436, 843.

\bibitem[\protect\citeauthoryear{Brown}{1995}]{Bro95}
Brown G.E. 1995, ApJ 440, 270.

\bibitem[\protect\citeauthoryear{Brown et al.}{2000}]{Bro00}
Brown G.E., Lee C.-H., Wijers R.A.M.J. Lee, H.K. Israelian G. and Bethe H.A. 2000, New Astronomy 5, 191.

\bibitem[\protect\citeauthoryear{Brown et al.}{2001}]{Bro01}
Brown, G.E., Heger, A., Langer, N., Lee, C.-H., Wellstein, S., and Bethe, H.A. 2001, New Astronomy 6, 457.

\bibitem[\protect\citeauthoryear{Brown et al.}{2007}]{Bro07}
Brown, G.E. Lee C.-H. and Moreno M\'endez E. 2007, ApJ 671, L41.

\bibitem[\protect\citeauthoryear{Brown et al.}{2008}]{BLMM08}   
Brown, G.E. Lee, C.-H. and Moreno M\'endez E. 2008, ApJ 685, 1063.

\bibitem[\protect\citeauthoryear{Chevalier}{1981,1989,1990}]{Che81}
Chevalier R.A. 1981, ApJ 246, 267.\\
-- 1989, ApJ 346, 847.\\
-- 1990, in SN 1987A and Other SNe, ed. J.J. Danziger (Elba:ESO)

\bibitem[\protect\citeauthoryear{Eggleton}{1983}]{Egg83}
Eggleton, P.P. 1983, {\it ApJ} 268, 368.

\bibitem[\protect\citeauthoryear{De Mink et al.}{2009}]{deM09}
De Mink S.E. Cantiello M., Langer N., Pols O.R., Brott, I. and Yoon S.-C., 2009, A\&A 497, 243.\\
-- 2008, IAUS 252, 365. \\
-- 2009, arXiv:0910.3694.

\bibitem[\protect\citeauthoryear{Gou et al.}{2009}]{Gou09}
Gou L., McClintock J.E., Liu J., Narayan R., Steiner J.F., Remillard R.A., Orosz J.A., and Davis S.W. 2009, arXiv:0901.0920v1.

\bibitem[\protect\citeauthoryear{Heger et al.}{2005}]{Heg05}
Heger, A., Woosley, S.E. and Spruit, H.C. 2005, ApJ 626, 350.

\bibitem[\protect\citeauthoryear{Hirschi, Meynet \& Maeder}{2005}]{Hir05}
Hirschi, R., Meynet, G. \& Maeder, A., A\&A 443, 581.

\bibitem[\protect\citeauthoryear{Lee et al.}{2000}]{Lee00}
Lee H.K., Wijers R.A.M.J. and Brown G.E. 2000, Physics Report 325,  83.

\bibitem[\protect\citeauthoryear{Lee et al.}{2002}]{Lee02}
Lee C.-H., Brown G.E., and Wijers R.A.M.J. 2002, ApJ 575, 996.

\bibitem[\protect\citeauthoryear{Liu et al.}{2008}]{Liu08}
Liu J., McClintock J.E., Narayan R., Davis S.W. and Orosz J. 2008, ApJ 679, L37.
--Erratum.  2010, ApJ 719, L109.

\bibitem[\protect\citeauthoryear{MacFadyen \& Woosley}{1999}]{Mac99}
MacFadyen, A.I. and Woosley, S.E. 1999, ApJ 524, 262.

\bibitem[\protect\citeauthoryear{Mohamed \& Podsiadlowski}{2007,2010}]{Moh07}
Mohamed, S. and Podsiadlowski, Ph., 2010, in preparation.\\
--2007, 15th European Workshop on White Dwarfs ASP Conf. Series, Vol. 372, p.397.

\bibitem[\protect\citeauthoryear{Moreno M\'endez et al.}{2007}]{Mor07}
Moreno M\'endez E., Brown G.E., Lee C.-H. \& Walter F. 2007, arXiv:astro-ph/0612461v4.

\bibitem[\protect\citeauthoryear{Moreno M\'endez et al.}{2008}]{Mor08}
Moreno M\'endez E., Brown G.E., Lee C.-H. \& Park I.H. 2008, ApJ 689, L9.

\bibitem[\protect\citeauthoryear{Mueller \& Vink}{2008}]{Mul08}
Mueller, P.E. \& Vink, J.S. 2008, {\it A\&A} 492, 493.

\bibitem[\protect\citeauthoryear{Orosz et al.}{2007}]{Oro07}
Orosz, J.A., McClintock, J.E., Narayan, R., Bailyn, C.D., Hartman J.D., Macri, L., Liu, J., Pietsch, W., Remillard, R.A., Shporer, A. \& Mazeh, T. 2007,
Nature 449, 872-875.

\bibitem[\protect\citeauthoryear{Orosz et al.}{2009}]{Oro09}
Orosz J.A., Steeghs D., McClintock J.E., Torres M.A.P., Bochkov I., Gou L., Narayan R., Blaschak M., Levine A.M., Remillard R.A., Bailyn C.D., Dwyer M.M. and Buxton M. 2009, ApJ 697, 573.

\bibitem[\protect\citeauthoryear{Podsiadlowski et al.}{1992}]{Pod92}
Podsiadlowski, Ph., Joss, P.C. and Hsu, J.J.L. 1992, ApJ 391, 246.

\bibitem[\protect\citeauthoryear{Sowers et al.}{1998}]{Sow98}
Sowers, J.W., Gies, D.R., Bagnuolo, Jr., W.G., Shafter, A.W.,
Wiemker, R., and Wiggs, M.S. 1998, ApJ 506, 424.

\bibitem[\protect\citeauthoryear{Spruit \& Phinney}{1998}]{Spr98}
Spruit, H.C. and Phinney, E.S. 1998, Nature 393, 139.

\bibitem[\protect\citeauthoryear{Steiner et al.}{2010}]{Ste10}
Steiner, J.F., Reis, C.R., McClintock, J.E., Narayan, R., Remillard, R.A., Orosz, J.A., Gou, L., Fabian, A.C. \& Torres, M.A.P. 2010, arXiv:1010.1013.

\bibitem[\protect\citeauthoryear{Valsecchi et al.}{2010}]{Val10}
Valsecchi, F., Glebbeeck E., Farr, W.M., Fragos, T., Willems, B., Orosz, J.A., Liu, J. \& Kalogera, V. 2010, arXiv:1010.4809.

\bibitem[Tauris \& van den Heuvel(2006)]{Tau06}
Tauris, T.M. \& van den Heuvel, E.P.J., 2006, Formation and evolution of compact stellar X-ray sources.
In {\it Compact stellar X-ray sources} (eds Lewin, W. and van der Klis, M.) 623-665 (Cambridge U.P., 2006).

\bibitem[\protect\citeauthoryear{Van den Heuvel}{1994}]{Heu94}
Van den Heuvel, E.P.J. 1994, Interacting Binaries: Topics In Close Binary Evolution. in {\it Lecture notes of the 22nd Advanced Course of the Swiss Society for Astronomy and Astrophysics (SSAA)} (eds Nussbaumer H. and  Orr A.) 263-474 (Berlin-Springer, 1994).

\bibitem[\protect\citeauthoryear{Van den Heuvel \& Yoon}{2007}]{vdH07}
Van den Heuvel E. and Yoon, S.-C. 2008, ApSS, 311, 177.

\bibitem[\protect\citeauthoryear{Yoon et al.}{2006}]{Yoo06}
Yoon S.-C., Langer N. \& Norman C. 2006, A\&A, 460, 199.

\bibitem[\protect\citeauthoryear{Zahn}{1977}]{Zah77}
Zahn, J.-P., 1977, A\&A, 57, 383.

\end{thebibliography}
\end{document}